\documentclass[anon,letterpaper,twocolumn,10pt]{article}
\RequirePackage[hyphens]{url}
\PassOptionsToPackage{hyphens}{url}\usepackage{hyperref}
\usepackage{usenix2019_v3}

\usepackage{verbatimbox}

\usepackage{amsmath}
\usepackage{amssymb}
\usepackage{booktabs}
\usepackage{caption}
\usepackage{color}
\usepackage{graphicx}
\usepackage{listings}
\usepackage{subcaption}
\usepackage{times}
\usepackage{comment}

\newcommand\like[1]{\begin{picture}(1,1)
\ifnum0=#1\put(.5,.35){\circle{1}}\else
\ifnum10=#1\put(.5,.35){\circle*{1}}\else
\put(.5,.35){\circle{1}}\put(.5,.35){\circle*{.#1}}
\fi\fi\end{picture}}

\newcommand{\systemname}{\textsc{Scalene}}
\newcommand{\systemcmd}{\texttt{scalene}}

\title{\systemname{}: Scripting-Language Aware Profiling for Python}

\author{
  {\rm Emery D. Berger}\\
  College of Information and Computer Sciences \\
  University of Massachusetts Amherst \\
  \texttt{emery@cs.umass.edu}
}

\newcommand{\punt}[1]{}

\punt{
\microtypesetup{
  final,
  tracking=true,
  kerning=true,
  spacing=true,
  factor=1100,
  stretch=20,
  shrink=20
}
\SetTracking{encoding={*}, shape=sc}{0} 
}

\definecolor{mygreen}{rgb}{0,0.6,0}
\definecolor{mygray}{rgb}{0.5,0.5,0.5}
\definecolor{mymauve}{rgb}{0.58,0,0.82}
\lstdefinelanguage{c++threads}[]{c++}{morekeywords={pthread_create,pthread_join,thread,join}}
\begin{document}

    \maketitle

    \begin{abstract}


Existing profilers for scripting languages (a.k.a. ``glue'' languages)
like Python suffer from numerous problems that drastically limit their
usefulness. They impose order-of-magnitude overheads, report
information at too coarse a granularity, or fail in the face of
threads. Worse, past profilers---essentially variants of their
counterparts for C---are oblivious to the fact that optimizing code in
scripting languages requires information about code spanning the
divide between the scripting language and libraries written in
compiled languages.

This paper introduces \emph{scripting-language aware profiling}, and
presents \systemname{},
an implementation of scripting-language aware profiling for
Python. \systemname{} employs a combination of sampling, inference,
and disassembly of bytecodes to efficiently and precisely attribute
execution time and memory usage to either Python, which developers can
optimize, or library code, which they cannot. It includes a novel
sampling memory allocator that reports line-level memory consumption
and trends with low overhead, helping developers reduce footprints and
identify leaks. Finally, it introduces a new metric, \emph{copy
volume}, to help developers root out insidious copying costs across
the Python/library boundary, which can drastically degrade
performance. \systemname{} works for single or multi-threaded Python
code, is precise, reporting detailed information at the line
granularity, while imposing modest overheads (26\%--53\%).

  \end{abstract}

  \section{Introduction}
  \label{sec:intro}

General-purpose programming languages can be thought of as spanning a
spectrum from systems languages to scripting
languages~\cite{ousterhout1998scripting}. Systems languages are
typically statically-typed and compiled, while scripting languages are
dynamically-typed and interpreted. As
Table~\ref{tab:scripting-languages} shows, scripting languages share
many implementation characteristics, such as unoptimized bytecode
interpreters, relatively inefficient garbage collectors, and limited
support for threads and signals~\footnote{We deliberately exclude
JavaScript, which was initially a scripting language; its
implementation has evolved to the point where it no longer has much in
common with those of other scripting languages, beyond its lack of support for threads.}.

\begin{table*}[!t]
    \centering
    \begin{tabular}{lccccccccccc}
        \textbf{Scripting} & \multicolumn{2}{c}{\textbf{Interpreter}} & \multicolumn{2}{c}{\textbf{GC algorithm}} & \multicolumn{2}{c}{\textbf{Threads}} & \multicolumn{2}{c}{\textbf{Signal limitations}} \\
        \textbf{Language}  & \textbf{Bytecode} & \textbf{AST}    & \textbf{Ref-counting} & \textbf{Mark-sweep} & \textbf{\texttt{pthreads}} & \textbf{Serialized}  & \textbf{Main Only} & \textbf{Delayed} \\
        \toprule
        \textbf{Perl} (1987)   &             & \checkmark & \checkmark &             &                     &              & \emph{N/A} & \checkmark \\
        \textbf{Tcl/Tk} (1988) & \checkmark  &            & \checkmark &             &                     &              & \emph{N/A} & \emph{N/A} \\
        \textbf{Python} (1990) & \checkmark  &            & \checkmark & $\dagger$   & \checkmark          & \checkmark   & \checkmark & \checkmark \\
        \textbf{Lua} (1993)    & \checkmark  &            &            & $\ddagger$  &                     &              & \checkmark & \checkmark \\
        \textbf{PHP} (1994)    & \checkmark  &            & $\dagger\dagger$ &        & *                  &              & \emph{N/A} & \checkmark \\
        \textbf{R} (1995)      & \checkmark  & \checkmark &            & \checkmark  &                     &              & \emph{N/A} & \emph{N/A} \\
        \textbf{Ruby} (1995)   & \checkmark  &            &            & $\ddagger\ddagger$  & \checkmark         & \checkmark   & \checkmark & \checkmark \\
    \end{tabular}
    \caption{\textbf{Major scripting language implementations share common implementation characteristics.} Next to each language is its first release date. All are dynamically typed; their standard implementations are interpreted and garbage-collected, most with reference counting. All lack threads or serialize them with a global interpreter lock (``GIL''), and all place severe limits on signal delivery, such as delivering only to the main thread and delaying delivery until the interpreter regains control (e.g., after executing a bytecode). $\dagger$: Python has an optional ``stop-the-world'' generational mark-sweep garbage collector. $\ddagger$: Lua garbage collector is an incremental mark-sweep collector. $\dagger\dagger$: PHP has a backup cycle collector~\cite{DBLP:conf/ecoop/BaconR01}. $\ddagger\ddagger$: Ruby's garbage collector is an incremental, generational mark-sweep collector. *: PHP support for threads is disabled by default, but can be configured at build time. ($\S\ref{sec:other-scripting-languages}$) \label{tab:scripting-languages}}
\end{table*}


This combination of overheads can lead applications in scripting
languages to run orders of magnitude slower than code written in
systems languages. They also can consume much more space: for example,
because of object metadata, an integer consumes 24--28 bytes in most
scripting languages. The widespread use of incomplete memory
management algorithms like reference counting, which cannot reclaim
cycles, only exacerbates the situation. These performance properties
combine to make developing efficient code in scripting languages a
challenge, but existing profilers for these languages are essentially
ports of profilers for systems languages
like \texttt{gprof}~\cite{DBLP:conf/sigplan/GrahamKM82}
or \texttt{perf}, which greatly limits their usefulness.

\punt{
For example, existing performance profilers for Python
like \texttt{cProfile}~\cite{cprofile} are largely ineffective at
helping programmers identify opportunities for performance
improvement. Their failure stems from multiple causes: first, they operate
at too coarse a granularity, focusing on functions and not on lines of
code.  Second, by focusing exclusively on code execution, they fail to
provide any way for programmers to reduce the insidious and invisible
overhead of data movement. These costs arise due to implicit copying
and serialization across the Python/C boundary, or accidental
instantiation of lazily generated objects.
}

\punt{
Scripting languages connect inefficient interpreters with
highly-efficient compiled libraries written in C or Java; optimizing
in scripting languages generally involves moving code into
libraries. Developers need to know if bottlenecks are in the
scripting language, which they can optimize, or in
libraries, which they cannot. Scripting languages can impose huge
per-object overheads (e.g., a float consumes 28 bytes in Python), and
often rely on reference counting garbage collectors, so developers
need to both limit unnecessary memory consumption and identify
leaks. Finally, developers need to eliminate implicit copying across
the scripting language/compiled language boundary, which can drastically degrade
performance.
}

This paper introduces \textbf{\emph{scripting-language aware
profiling}}, which directly addresses the key challenges of optimizing
code written in scripting languages. Because scripting languages are
so inefficient, optimizing applications in these languages generally
involves moving code into native libraries. Developers thus need to
know if bottlenecks reside in the scripting language, which they can
optimize, or in native libraries, which they cannot. Because of the
significant space overheads that scripting languages impose,
developers need to both limit unnecessary memory consumption by
avoiding accidental instantiation of lazily generated objects, moving
memory intensive code into libraries, as well as identify
leaks. Finally, they need to identify and eliminate implicit copying
across the scripting language/compiled language boundary, which can
drastically degrade performance.

We have developed a scripting-language aware profiler for
Python called \textbf{\systemname{}}. We target Python because it is 
one of the most popular scripting languages according to a variety of
rankings~\cite{ieeeplrank2019,redmonk-rankings,tiobe-index,stack-overflow-2019-survey}.
Large-scale industrial users of Python include
Dropbox~\cite{python-at-dropbox},
Facebook~\cite{python-at-facebook}, 
Instagram~\cite{python-at-instagram},
Netflix~\cite{python-at-netflix}, Spotify~\cite{python-at-spotify},
and YouTube~\cite{python-at-youtube}.

In addition to subsuming the functionality of previous profilers with
higher performance, \systemname{} implements the following novel
scripting-aware profiling features:

\begin{table*}[!t]
    \centering
    \begin{tabular}{lccccccccc}
        \textbf{Profiler} & \textbf{Time} & \textbf{Efficiency} & \textbf{Mem} & \textbf{Unmodified} & \textbf{Threads} & \multicolumn{3}{c}{\textbf{\emph{Scripting-Lang Aware}}} \\
                          & & & \textbf{Cons.} & \textbf{Code} & & \textbf{Python/C} & \textbf{Mem Trend} & \textbf{Copy Vol.} \\ \toprule
        \emph{function-granularity} \\ 
        \texttt{cProfile}~\cite{cprofile}  & real  & \like{10} &           & \checkmark &  & \\ 
        \texttt{Profile}~\cite{profile}    & CPU         & \like{3}  &           & \checkmark &  & \\ 
        \texttt{pyinstrument}~\cite{pyinstrument}  & real  & \like{9}  &           & \checkmark &  & \\ 
        \texttt{yappi}$_\emph{\mbox{CPU}}$~\cite{yappi} & CPU         & \like{4}  &           & \checkmark & \checkmark & \\ 
        \texttt{yappi}$_\emph{wallclock}$~\cite{yappi}  & real  & \like{8}  &           & \checkmark & \checkmark & \\
        \midrule
        \emph{line-granularity} \\
        \texttt{line\_profiler}~\cite{line_profiler}    & real  & \like{5}  &           &            & & \\ 
        \texttt{pprofile}$_\emph{\mbox{det}}$~\cite{pprofile}
                                           & real  & \like{2}  &           & \checkmark & \checkmark & \\ 
        \texttt{pprofile}$_\emph{\mbox{stat}}$~\cite{pprofile}
                                           & real  & \like{10}  &           & \checkmark & \checkmark & \\ 
        \texttt{py-spy}~\cite{py_spy}      & \textbf{both}& \like{10} &           & \checkmark & \checkmark & \\ 
        \texttt{memory\_profiler}~\cite{memory_profiler} & \emph{N/A}  & \like{1}  & \checkmark &     &  & \\
        \midrule
        \textbf{\systemname{}} & \textbf{both} & \like{9} & \checkmark & \checkmark & \checkmark & \checkmark & \checkmark & \checkmark \\
    \end{tabular}
    
    \caption{\textbf{Existing Python profilers vs. \systemname{}.}
      \emph{Time} indicates real (wall-clock) time, CPU time, or
      both. Darker circles shown in \emph{Efficiency} indicate higher
      efficiency (lower overheads), ranging from less than $1.2\times$
      to over $1000\times$ (Figure~\ref{fig:profiler-overheads}
      provides detailed performance breakdowns, and
      Section~\ref{sec:python-profilers} provides other details.)
      \emph{Mem Cons.} indicates whether it profiles memory
      consumption. \emph{Unmodified Code} means that use of the
      profiler does not require source code
      modifications. \emph{Threads} indicates whether it correctly
      attributes execution time or memory consumption for
      multithreaded Python code. Only \systemname{} reports
      scripting-language aware statistics: \emph{Python/C} = separate
      attribution of execution time ($\S\ref{sec:python-versus-c}$)
      and memory ($\S\ref{sec:call-stack-sampling}$) to Python code
      or C, \emph{Mem Trend} = timeline of memory consumption
      ($\S\ref{sec:memory-trends}$), and \emph{Copy Vol.} = \emph{copy
        volume} in MB/s
      ($\S\ref{sec:copy-volume}$).\label{tab:comparison}}
    
\end{table*}

\begin{itemize}

\item \textbf{Separate Python/C accounting of time and space.}
\systemname{} separately attributes both execution time 
($\S\ref{sec:python-versus-c}$) and memory consumption
($\S\ref{sec:call-stack-sampling}$) based on whether it stems from
Python or native code. Most Python programmers are
not able to optimize the performance or memory consumption of native
code (which is usually either in the Python implementation or external
libraries), so this helps developers focus their optimization efforts
on the code they can improve.

\item \textbf{Fine-grained tracking of memory use over time.} \systemname{} uses a novel
\emph{sampling memory allocator} ($\S\ref{sec:memory-usage}$) to not only enable separate
accounting of memory consumption to Python vs. native code, but also
to efficiently profile memory usage at the line granularity. It
produces \emph{per-line memory profiles} in the form of sparklines
(see Figure~\ref{fig:example}): these are in-line graphs that indicate
trends of memory consumption over time, making it easier to track down
leaks ($\S\ref{sec:memory-trends}$).

\item \textbf{Copy volume.} Finally, \systemname{} reports copy volume in megabytes per second,
for each line of code ($\S\ref{sec:copy-volume}$). This novel metric
makes it straightforward to spot inadvertent copying, including silent
coercion or crossing the Python/library boundary (e.g., accidentally
converting \texttt{numpy} arrays into Python arrays or vice versa).

\end{itemize}

\systemname{} overcomes a number of technical challenges inherent to
the implementation of scripting languages to collect this information
with relatively low performance overhead. \systemname{} outperforms
other profilers by in some cases orders of magnitude, while delivering
far more detailed information. \systemname{} is \emph{precise}. Unlike
many existing Python profilers, \systemname{} performs both memory and
CPU profiling \emph{at the line granularity}. This level of detail can
be much more useful than the function-granularity profiles returned by
many profilers: unlike in systems languages, where individual lines
are often compiled to a few cycles, lines of code in scripting
languages are often orders of magnitude more expensive. Our prototype
achieves this precision with low overhead. For full memory and copy
profiling, it imposes between 26\%--53\% overhead; for CPU profiling
only (separating Python and C execution), it imposes no observable
performance penalty (Section~\ref{sec:evaluation}).

While this paper primarily focuses on \systemname{} and Python, we
believe the techniques it describes depend primarily on
implementation details common to almost all scripting languages, and
thus should be broadly applicable.

    \section{Overview of \systemname{}}
    \label{sec:overview}

\begin{figure*}[!t]
  \begin{subfigure}[t]{0.99\linewidth} \includegraphics[width=\textwidth]{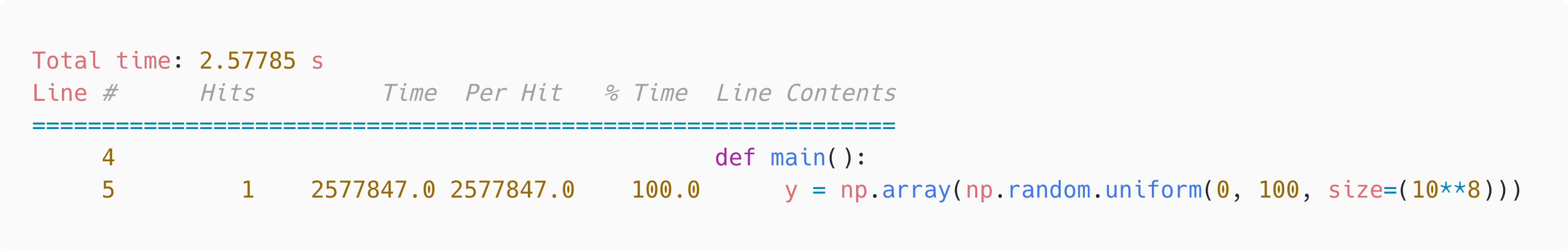} \caption{\textbf{Profiling
    with \texttt{line\_profiler}}. Traditional CPU profilers often
    yield little actionable
    insight.\label{fig:line-profiler-example}} \vspace{1em} \end{subfigure} \begin{subfigure}[t]{0.99\linewidth} \includegraphics[width=\textwidth]{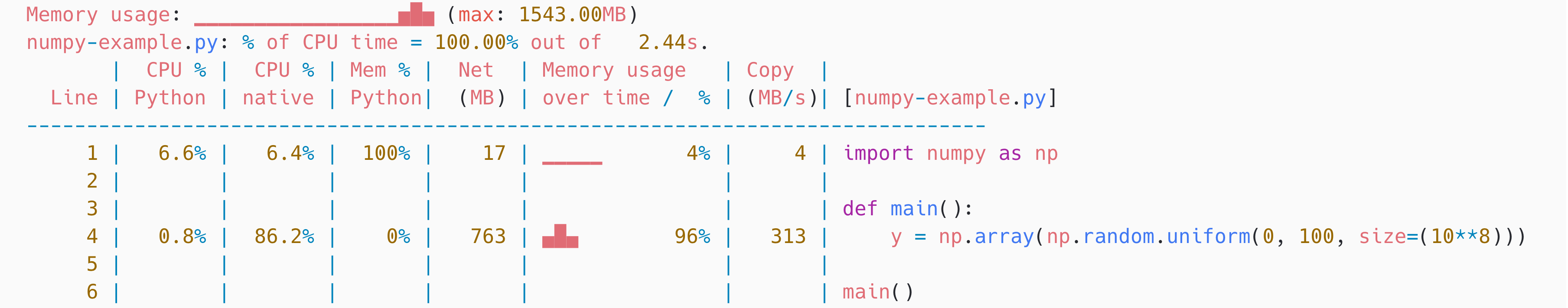} \caption{\textbf{Profiling
    with \systemname{}: before optimization.} Line 4's sawtooth
    allocation and high copy volume indicate copying due
    to \texttt{np.array}.\label{fig:example-before}} \vspace{1em} \end{subfigure} \begin{subfigure}[t]{0.99\linewidth} \includegraphics[width=\textwidth]{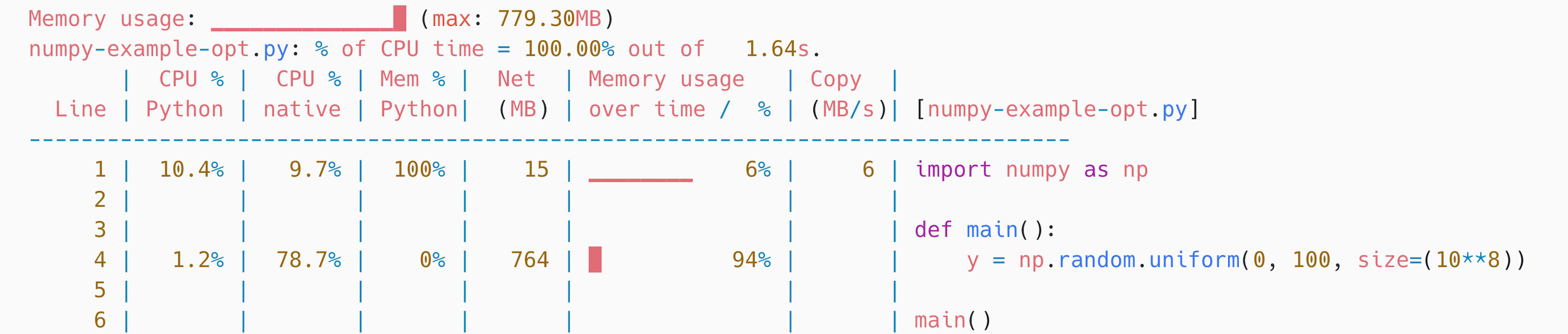} \caption{\textbf{Profiling
    with \systemname{}: after optimization.}  Removing the call
    to \texttt{np.array} cuts execution time and total memory
    footprint in
    half.\label{fig:example-after}} \vspace{1em} \end{subfigure} \caption{\textbf{\systemname{}'s
    profiler can effectively guide optimization efforts.} Unlike past
    profilers, \systemname{} splits time spent and memory consumed in
    the Python interpreter vs. native libraries, includes average net memory
    consumption as well as memory usage over time, and reports copy
    volume. The sawtooth pattern and high copy volume on line 4 in
    Figure~\ref{fig:example-before} indicate unnecessary allocation
    and copying due to a redundant \texttt{np.array} call. Removing it
    stabilizes allocation and eliminates copying overhead, leading to
    a 50\% performance improvement and footprint
    reduction.\label{fig:example}}
\end{figure*}

This section provides an overview of \systemname{}'s operation in
collecting profile information.

Profiling a Python program with \systemname{} is a straightforward
matter of replacing the call to Python (e.g., \texttt{python3 app.py}
becomes \texttt{\systemcmd{} app.py}). By default, \systemname{}
generates a profile when the program terminates. To support
long-running Python applications, \systemname{} also can be directed
via command-line parameters to periodically write profiles to a file.

In addition to providing line-granularity CPU profiles, \systemname{}
breaks out CPU usage by whether it is attributable to interpreted or
native code ($\S\ref{sec:python-versus-c}$). Its sampling memory
allocator---which replaces the default allocator through library
interposition---lets it report line-granularity net memory
consumption, separately attribute memory consumption to Python or
native code ($\S\ref{sec:memory-usage}$), and display trends, in the form
of ``sparklines''~\cite{tufte2006beautiful}, which capture memory
usage over time ($\S\ref{sec:memory-trends}$). This information makes
it easy for developers to identify leaks or unnecessary allocation and
freeing. It also reports copy volume in megabytes per second, which
can identify unnecessary copying, whether in Python, in native
libraries, or across the boundary.

Figure~\ref{fig:example} demonstrates how \systemname{}'s guidance can
help developers find inefficiencies and optimize their
code. Figure~\ref{fig:line-profiler-example} shows a profile from a
standard Python profiler, \texttt{line\_profiler}. The generic nature
of past profilers (just tracking CPU time) often fails to yield
meaningful insights. Here, it indicates that the line of code is
responsible for 100\% of program execution, but this fact does not
suggest optimization opportunities.

By contrast, Figure~\ref{fig:example-before} shows the output of
\systemname{} for the same program. The profile reveals that the line of code
in question is unusual: its memory consumption (exclusively in native
code) exhibits a distinctive sawtooth pattern. In addition, the line
is responsible for a considerable amount of copy volume (almost 600
MB/s). Together, this information tells a familiar tale: copying to a
temporary, which is allocated and then promptly discarded. Inspection
of this line of code reveals an unnecessary call to \texttt{np.array}
(the result of the expression is already a \texttt{numpy}
array). Removing that call, as Figure~\ref{fig:example-after} shows,
reduces both overall memory consumption (shown in the top line of the
profile) and total execution time by 50\%.

In addition to revealing optimization opportunities that other
profilers cannot, \systemname{} is also fast, imposing just 10\%
overhead for this benchmark. The next section details
how \systemname{}'s implementation simultaneously delivers high
precision and generally low overhead (at most 53\%).

    \section{Implementation}
    \label{sec:implementation}

\begin{figure}[!t]
\centering
  \includegraphics[width=.8\linewidth]{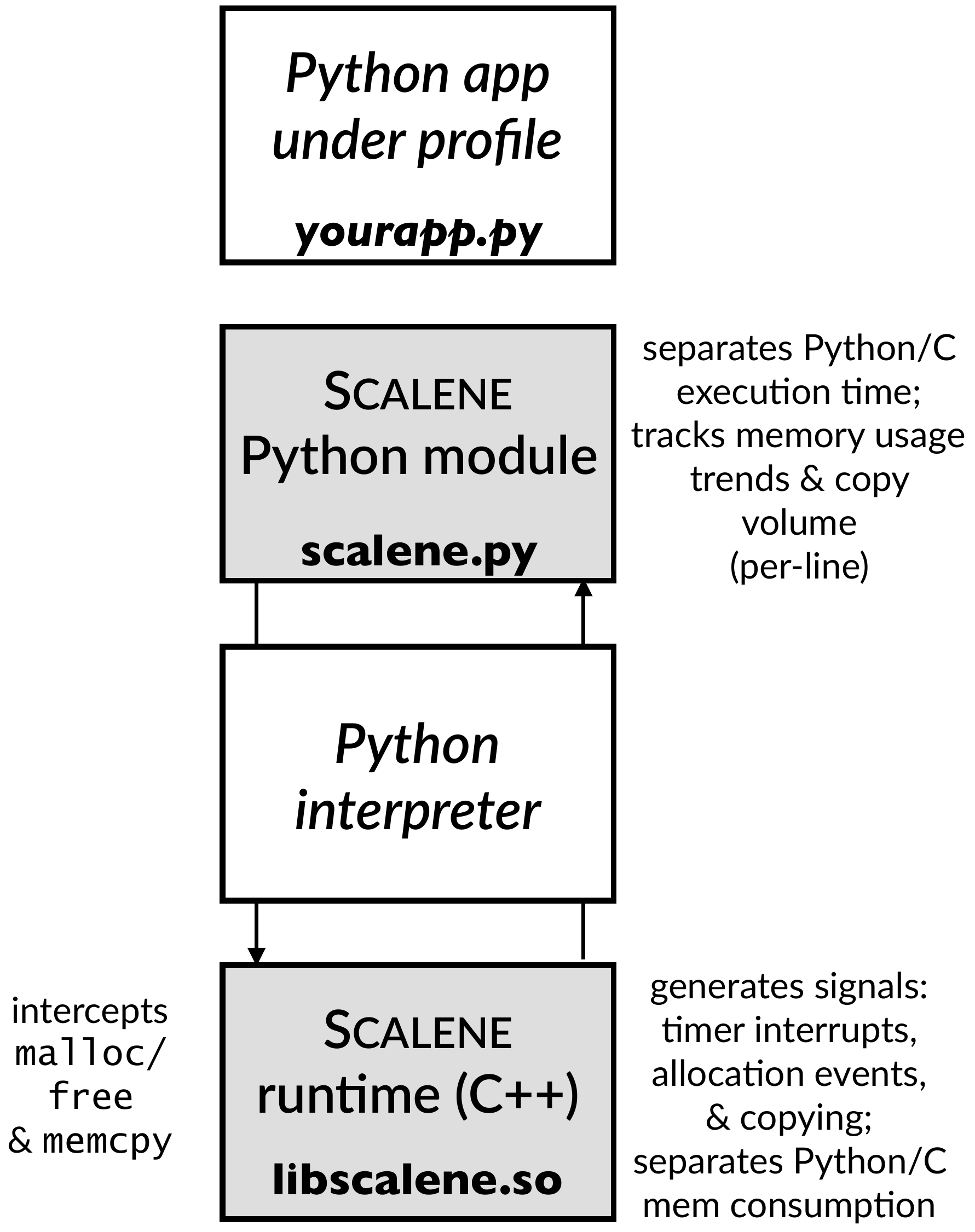}
  \caption{\textbf{\systemname{} Overview.} \systemname{} consists of two main components, a Python module and a C++-based runtime system, both depicted in gray. The runtime system is loaded via library interposition. The white components (the code being profiled and the Python interpreter itself) require no modifications.\label{fig:overview}}
  \vspace{1em}
\end{figure}

Our \systemname{} prototype runs on Linux (including Windows Subsystem
for Linux, version 2) and Mac OS X, for Python versions 3.5 and
higher. It is implemented as a combination of a pure Python module and
a specialized runtime library written in C++ that replaces key calls
by library interposition (that is, \texttt{LD\_PRELOAD} on Linux
and \texttt{DYLD\_INSERT\_LIBRARIES} on Mac OS
X). Figure~\ref{fig:overview} presents a diagrammatic overview.

Crucially, \systemname{} does not depend on any modifications to the
underlying CPython interpreter. This approach means that \systemname{}
works unchanged with other implementations of Python like
PyPy~\cite{DBLP:conf/oopsla/RigoP06}. It also provides evidence that
the techniques we develop for \systemname{} should be portable to
other scripting languages without significant
changes. Table~\ref{tab:scripting-language-features} presents an
overview of scripting languages and the features that \systemname{}
relies on.

Exposing scripting-language aware features---without modifying the
underlying language---required overcoming a number of technical
challenges.  This section first explains how \systemname{} turns the
severe limitations on signal delivery (typical of scripting languages)
to good effect. It then presents \systemname{}'s runtime library,
which cooperates with the Python-based component to track memory
usage, trends over time, and copy volume, all at a line granularity
and with low overhead. In the remainder of this section, we focus our
discussion specifically on Python, noting where
characteristics of Python differ from other scripting languages.

\subsection{Python/C Attribution of CPU Time}
\label{sec:python-versus-c}

Traditional sampling profilers work by periodically interrupting
program execution and examining the current program counter. Given a
sufficiently large number of samples, the number of samples each
program counter receives is proportional to the amount of time that
the program was executing. Sampling can be triggered by the passage of
real (wall-clock) time, which accounts for CPU time as well as time
spent waiting for I/O or other events, or virtual time (the time the
application was actually scheduled for execution), which only accounts
for CPU time.

While both timer approaches are available in Python (on Linux and Mac
OS X systems), directly using sampling is ineffective for Python.  As
noted previously, nearly all scripting languages impose severe
limitations on signal delivery. Typically, as in Python, these signals
are delayed until the virtual machine (i.e., the interpreter loop)
regains control, often after each opcode. These signals are also only
delivered to the main thread.

The result is that \emph{no} signals are delivered---and thus, no
samples accrue---during the entire time that Python spends executing
external library calls. It also means that lines of code executing in
threads (besides the main thread) are never executed. In the worst
case, sampling can utterly fail. Consider a main thread that spawns
child threads and then blocks waiting for them to finish. Because no
signals are delivered to the main thread while it is blocking, and
because the threads themselves also never receive signals, a na\"ive
sampling profiler could report that no time elapsed. (Note that because
of serialization due to the GIL, Python threads are not
particularly well suited for parallel code, but they are widely used
in servers to manage connections.)

\subsubsection*{Inferring Time Spent in C Code}

Recall that one of the goals of \systemname{} is to attribute
execution time separately, so that developers can identify which code
they can optimize (Python code) and which code they generally cannot
(C or other native code). An apparently promising approach would be
handle signals, walk the stack, and distinguish whether the code was
invoked by the interpreter as an external function, or whether it was
within the interpreter itself. However, as we note above, no signals
are delivered during native code execution, making such an approach
impossible.

Instead, we turn this ostensible limitation to our advantage. We
leverage the following insight: \emph{any delay in signal delivery
corresponds to time spent executing outside the interpreter}. That is,
if \systemname{}'s signal handler received the signal immediately
(that is, in the requested timing interval), then all of that time
must have been spent in the interpreter. If it was delayed, it must be
due to running code outside the interpreter, which is the only cause
of delays (at least, in virtual time).

To track this time, \systemname{} uses a clock
(either \texttt{time.process\_time()}
or \texttt{time.perf\_counter()}) to record the last time it received a
CPU timer interrupt. When it receives the next interrupt, it computes
$T$, the elapsed time and compares it to the timing interval $q$ (for quantum).

\systemname{} uses the following algorithm to assign time to Python or C:
Whenever \systemname{} receives a signal, \systemname{} walks the
Python stack until it reaches code being profiled (that is, outside of
libraries or the Python interpreter itself), and attributes time to
the resulting line of code. \systemname{} maintains two counters for
every line of code being profiled: one for Python, and one for C
(native) code. Each time a line is interrupted by a
signal, \systemname{} increments the Python counter by $q$, the timing
interval, and it increments the C counter by $T-q$.

It might seem counterintuitive to update both counters, but as we show
below, this approach yields an unbiased estimator. That is, in
expectation, the estimates are equivalent to the actual execution
times.  We first justify this intuitively, and then formally prove
it is unbiased.

First, consider a line of code that spends 100\% of its time in the
Python interpreter. Whenever a signal occurs during execution of that
line, it will be almost immediately delivered, meaning that $T =
q$. Thus, all of its samples ($q$) will accumulate for the Python
counter, and 0\% ($T - q = T - T = 0$) for the C counter, yielding an
accurate estimate.

Now consider a line that spends 100\% of its time executing C
code. During that time, no signals are delivered. The longer the time
elapsed, the more accurate this estimate becomes. The ratio of time
attributed to C code over (C plus Python) is $\frac{T-q}{(T-q)+q}$,
which simplifies to $\frac{T-q}{T}$.  As $T$ approaches infinity, this
expression approaches 1 (that is,
$\lim_{T\rightarrow\infty}\frac{T-q}{T} = 1$), making it an accurate
estimate.

While equality holds in the limit, the resulting approximation is
accurate even for relatively low elapsed times, as long as they are
larger relative to the sampling interval. \systemname{}'s current
sampling interval is 0.01 seconds, so a line that takes one second
executing C code would receive $(1-0.01)/1$ or 99\% of its samples as
native code, which is only off by 1\%.

Finally, consider the general case when the ratio of time spent in C
code to Python code is some fraction $F$. In expectation, the signal
will be delayed with probability $F$, meaning that the attribution to
C code will be $\frac{F(T-q)}{T}$. As $T$ approaches infinity, this
expression approaches $F$.

To prove that this approach yields an unbiased estimator, we need to
show that, in expectation, the estimates equal the actual values.  We
denote the execution time of the program's Python and C
components as $P$ and $C$, respectively. We subscript these with an
index (e.g., $P_i$) to denote individual lines of code; $P
= \sum_i{P_i}$. We use hats to denote estimates, as in $\hat{P_i}$.
Rephrasing formally, to show that these estimates are unbiased, we
need to show that $E[\hat{P_i}] = E[P_i]$ and $E[\hat{C_i}] = E[C_i]$.

We first observe that, in expectation, $P_i$ is the proportional
fraction of execution time taken by line $i$ of the total $P$ (and
similarly for $C_i$). By linearity of expectation, it is sufficient to
consider the total execution times and show that $E[\hat{P}] = E[P]$
and $E[\hat{C}] = E[C]$.

Call $S$ the total number of samples received by the program---by
definition, only when it is executing Python code. This means that
$E[P] = Sq$: the expected running time of Python code is the number of
samples times the length of each quantum.
\systemname{} adds $q$ every time to its estimate of Python execution time whenever it receives
a signal: the total is $Sq$, so $E[\hat{P}] = Sq = E[P]$. For C
code, \systemname{} adds the time elapsed waiting for a signal. The
total time elapsed when waiting for a signal is the total elapsed time
minus the time accounted for by signals: $E[\hat{C}] = E[(P + C) -
Sq]$. We have already shown that $E[P] = Sq$, so we have $E[\hat{C}] =
E[C]$.

\paragraph{Simulation study.} To quantify how quickly these formulas converge
converges depending on the ratio of $T$ and $q$, we perform a
simulation study.  The simulator mimics the effect of executing a
Python program line by line, spending a random amount of time
executing Python code, and then a random amount of time running C
code. The simulator draws the execution times of the Python and C
components of each line of code from a Pareto distribution such that
20\% of the code accounts for 80\% of the total execution time
($\alpha = 1.16$). It then simulates execution of 100 lines of code
for a range of execution times, where the simulated quantum is set at
0.01 seconds (as in \systemname{}), and attributes time as described
either to Python or C code. At the end of execution, the simulator
reports the estimated total time spent in Python code or C code, along
with the simulated ``actual'' time.

\begin{figure}[!t]
\centering
  \includegraphics[width=.8\linewidth]{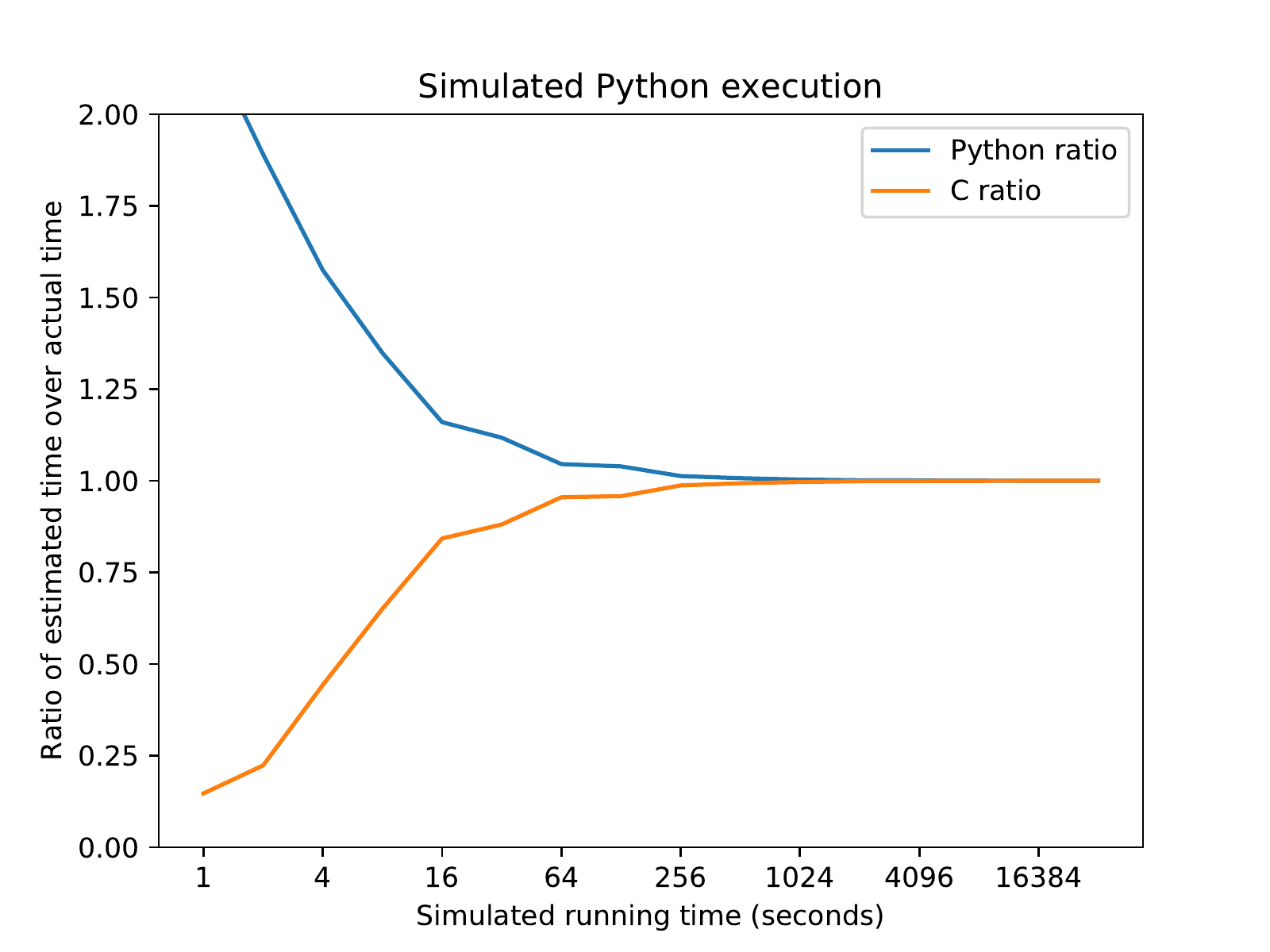}
  \caption{\textbf{Simulated execution of Python/C code.} This graph validates \systemname{}'s inference approach to distinguishing Python and C execution times, showing that as execution time increases, the estimated shares of execution time become increasingly accurate.\label{fig:simulation}}
  \vspace{1em}
\end{figure}

Figure~\ref{fig:simulation} presents the results of running this
simulation 10 times; the x-axis is execution time, and the y-axis is
the average ratio of estimated time to simulated time. As predicted,
the accuracy of both estimators increases as execution time
increases. The simulation shows that the amount of error in both
estimates is under 10\% after one minute of execution
time. Empirically, we find that actual code converges more quickly; we
attribute this to the fact that actual Python code does not consist of
serialized phases of Python and then C code, but rather that the
phases are effectively randomly mixed.

We also evaluate the correlation of all estimated and simulated times
using Spearman's $\rho$, which measures whether there is a linear
relationship between the two, a value of $\rho=1$ denoting a monotonic
linear relationship between the values. For 64 seconds of execution,
the correlation coefficient for the Python estimates and the C
estimates to their simulated execution time is $\rho > 0.99$,
indicating that the estimates are directly correlated with the
simulated times.


\subsubsection*{Attributing Time Spent in Threads}

The approach described above accurately attributes execution time for
Python vs. C code in the main thread, but it does not attribute
execution time at all for threads, which themselves never receive
signals. To handle this, \systemname{} relies on the following Python
features, which are available in other scripting
languages: \emph{monkey patching}, \emph{thread
enumeration}, \emph{stack inspection}, and \emph{bytecode
disassembly}.


\paragraph{\textbf{Monkey patching.}}
Monkey patching refers to the redefinition of functions at runtime, a
feature of most scripting languages. \systemname{} uses monkey
patching to ensure that signals are always received by the main
thread, even when that thread is blocking. Essentially, it replaces
blocking functions like \texttt{threading.join} with ones that always
use timeouts. The timeout interval is currently set to Python's thread
quantum (obtained via \texttt{sys.getswitchinterval()}). By replacing these calls,
\systemname{} ensures that the main thread yields frequently, allowing
signals to be delivered regularly.

In addition, to attribute execution times correctly, \systemname{}
maintains a status flag for every thread, all
initially \emph{executing}. In each of the calls it intercepts,
before \systemname{} actually issues the blocking call, it sets the
calling thread's status as \emph{sleeping}.  Once that thread returns
(either after successfully acquiring the desired resource or after a
timeout), \systemname{} resets the status of the calling thread to
executing. \systemname{} only attributes time to currently executing
threads.

\paragraph{\textbf{Thread enumeration.}}
When the main thread receives a signal, \systemname{} introspects on
all running threads, invoking \texttt{threading.enumerate()} to
collect a list of all running threads; similar logic exists in other
scripting languages (see Table~\ref{tab:scripting-language-features}).

\paragraph{\textbf{Stack inspection.}}
\systemname{} next obtains the Python stack frame from each thread using
Python's \texttt{sys.\_current\_frames()} method. Note that the
preceding underscore is just Python convention for a ``protected''
class method or variable. As above, \systemname{} walks the stack to
find the appropriate line of code for which it will attribute
execution time.

\paragraph{\textbf{Bytecode disassembly.}}
Finally, \systemname{} uses bytecode disassembly (via the \texttt{dis}
module) to distinguish between time spent in Python vs. C
code. Whenever Python invokes an external function, it does so via a
bytecode whose textual representation starts with \texttt{CALL\_}
(this approach is common to other languages; for example, Lua
uses \texttt{OP\_CALL}, while Ruby's is
\texttt{opt\_call\_c\_function}). \systemname{} builds a map of
all such bytecodes at startup.

For each running thread, \systemname{} checks the stack and its
associated map to determine if the currently executing bytecode is a
call instruction. Because this method lets \systemname{} know with
certainty whether the thread is currently executing Python or C code,
there is no need for the inference algorithm described above. If the
bytecode is a call, \systemname{} assigns time to the C counter;
otherwise, it assigns it to the Python counter.


\subsection{Memory Usage}
\label{sec:memory-usage}

Traditional profilers either report CPU time or memory
consumption; \systemname{} reports both, at a line granularity.  It is
vital that \systemname{} track memory both inside Python and out;
external libraries are often responsible for a considerable fraction
of memory consumption.

To do this, \systemname{} intercepts all memory allocation related
calls (\texttt{malloc}, \texttt{free}, etc.) via its own replacement
memory allocator, which is injected before execution begins.

By default, Python relies on its own internal memory allocator for
objects 512 bytes or smaller, maintaining a freelist of objects for
every multiple of 8 bytes in size. However, if the environment
variable \texttt{PYTHONMALLOC} is set to \texttt{malloc}, Python will
instead use \texttt{malloc} to satisfy all object
requests. \systemname{} sets this variable accordingly before
beginning profiling. Note that some other languages may not make it so
straightforward to replace all allocations; for example, while Ruby
uses the system \texttt{malloc} to satisfy large object requests,
there is no facility for replacing small object allocations. However,
most other scripting languages make it simple to redirect all of their
allocations (see Table~\ref{tab:scripting-language-features}).

\subsubsection*{An Efficient Replacement Allocator}

Because Python applications can be extremely allocation-intensive,
using a standard system allocator for all objects can impose
considerable overhead.  In our experiments, replacing the allocator by
the default on Mac OS X can slow down execution by 80\%.  We viewed
this as an unacceptably large amount of overhead, and ended up
building a new allocator in C++, with some components drawn from the
Heap Layers infrastructure~\cite{DBLP:conf/pldi/BergerZM01}.

This might at first glance seem unnecessary, since in theory, one
could extract the allocator from the Python source code and convert it
into a general-purpose allocator.  Unfortunately, the existing Python
allocator is also not suitable for use as a general \texttt{malloc}
replacement. First, the built-in Python allocator is implemented on
top of \texttt{malloc}; in effect, making it a general-purpose
allocator still would require building an implementation
of \texttt{malloc}.

However, the most important consideration, which necessitates a
redesign of the algorithm, is that a usable general-purpose allocator
replacement needs to be robust to invocations of \texttt{free} on
foreign objects. That is, it must reject attempts to free objects
which were not obtained via calls to its \texttt{malloc}. This case is
not a theoretical concern, but is in fact a near certitude.  It can
arise not just due to programmer error (e.g., freeing an unaligned
object, a stack-allocated object, or an object obtained from an
internal allocator), but also because of timing: library interposition
does not necessarily intercept all object allocations. In fact, Python
invokes \texttt{free} on ten foreign objects, which are allocated
before \systemname{}'s interposition completes. Because re-using
foreign objects to satisfy object requests could lead to havoc, a
general-purpose allocator needs a fast way to identify foreign objects
and discard them (a small leak being preferable to a crash).

We therefore built a general-purpose memory allocator
for \systemname{} whose performance characteristics nearly match those
of the Python allocator. At initialization, the \systemname{}
allocator reserves a contiguous range of virtual memory to satisfy
small object requests. It also allocates memory for large objects to
be aligned to 4K boundaries, and places a magic number
(\texttt{0xDEADBEEF}) in each header as a validity check. If objects
are outside the contiguous range, not properly aligned, or fail their
validity check,
\systemname{} treats them as foreign. We have found this approach to be sufficiently robust to enable it to
work on every Python program we have tested.

Otherwise, the internals of the \systemname{} allocator are similar in
spirit to those of the Python allocator; it maintains lists for every
size class of a multiple of 16 bytes up to 512 bytes. These point to
4K slabs of memory, with a highly optimized allocation fast
path. Large objects are allocated separately, either from a store of
4K chunks, or directly via \texttt{mmap}. In our tests, this allocator
significantly closes the performance gap between the system allocator
and Python's internal allocator, reducing overhead from 80\% to around
20\%. We expect to be able to optimize performance further, especially
by avoiding repeated calls to \texttt{mmap} for large object allocation.

\subsubsection*{Sampling}

With this efficient allocator in hand intercepting all allocation
requests, we are now in a position to add the key component: sampling.

\paragraph{Allocation-Triggered Sampling:} The \systemname{} sampling allocator maintains a count of all memory
allocations and frees, in bytes. Once either of these crosses a
threshold, it sends a signal to the Python process. To
allow \systemname{} to work on Mac OS X, which does not implement
POSIX real-time signals, we re-purpose two rarely used signals:
\texttt{SIGXCPU} for \texttt{malloc} signals,
and \texttt{SIGXFSZ} for \texttt{free} signals. \systemname{} triggers these
signals roughly after a fixed amount allocation or freeing.
This interval is currently set as a prime number above $1MB$, intended
to reduce the risk of stride behavior interfering with sampling.

\paragraph{Call Stack Sampling:}
\label{sec:call-stack-sampling}

To track the provenance of allocated objects (that is, whether they
were allocated by Python or native code), \systemname{} triggers call
stack sampling. The sampling rate is set as a multiple of the
frequency of allocation samples (currently $13\times$). Whenever the
threshold number of allocations is crossed (that is, after $1{MB}/13$
allocations), \systemname{} climbs the stack to determine whether the
sampled allocation came from Python or native code.

To distinguish between these two, \systemname{} relies on the
following domain-specific knowledge of Python internals. Python has a
wide range of functions that create new Python references, all of
which begin with either \texttt{Py\_}
or \texttt{\_Py}. If \systemname{} encounters one of these functions
as it climbs the stack, the object was by definition allocated by
Python, so it increments a count of Python allocations by the
requested size.\footnote{A few special cases: \texttt{\_PyCFunction}
allocates memory but on behalf of a C call, and \texttt{PyArray}, a
non-Python call that \texttt{numpy} uses for allocating its own
(native) arrays; \systemname{} treats both of these correctly as C
allocations.}

After walking a maximum number of frames (currently
4), if \systemname{} has not encountered one of these functions, it
concludes that the allocation was due to native code and increments the
C allocation counter. When the \systemname{} allocator eventually sends
allocation information to the Python module (described below), it
includes the ratio of Python bytes over total allocated bytes. It
then resets both allocation counters.

Because resolving function names via \texttt{dladdr} is relatively
costly, especially on Mac OS X, \systemname{} maintains an
open-addressed hash table that maps call stack addresses to function
names. This hash table is a key optimization: using it
reduces \systemname{}'s overhead by 16\% in one of our benchmarks.

\paragraph{Managing Signals:}
Because Python does not queue signals, signals can be lost. We thus
need a separate channel to communicate with the main process; to do
this, we allocate a temporary file with the process-id as a
suffix. \systemname{} appends information about allocations or frees
to this file, as well as the fraction of Python allocations.

When \systemname{}'s signal handler is triggered (in the Python module),
it reads the temporary file and attributes allocations or
frees to the currently executing line of code in every frame. As
with sampling CPU execution, lines of code that frequently allocate or
free memory will get more samples. \systemname{} also tracks the
current memory footprint, which it uses both to report maximum memory
consumption and to generate sparklines for memory allocation
trends (Section~\ref{sec:memory-trends}).

One fly in the ointment is that the Python signal handler itself
allocates memory. Unlike in C, this allocation is impossible to avoid
because the interpreter itself is constantly allocating and freeing
memory. However, \systemname{} again leverages one of Python's
limitations to its advantage: Python's global interpreter lock ensures
that there is no true concurrency inside the interpreter.
Therefore, \systemname{} straightforwardly prevents re-entrant
calls by checking a boolean flag to see if it is already in the signal
handler; if not, it sets the flag.

\subsection{Memory Trends}
\label{sec:memory-trends}

\systemname{} not only reports net memory consumption per line, but
also reports memory usage over time in the form of sparklines, both
for the program as a whole and for each individual line. It adds the
current footprint (updated on every allocation and free event,
comprising at least $1MB$ of allocation) to an ordered array of
samples for each line of code. The sampling array is chosen to be a
multiple of 3, currently 27. When the array fills, the contents are
reduced by a factor of 3, replacing each entry by its median value;
after this reduction, footprint samples are again added to the end of
the array. The effect of this approach is to smooth older footprint
trends (on the left side of the sparkline) while maintaining higher
fidelity for more recent footprints.

\subsection{Copy Volume}
\label{sec:copy-volume}

Finally, \systemname{} reports copy volume by line. It also
accomplishes this by sampling. The \systemname{} runtime system
interposes on \texttt{memcpy}, which is invoked both for general
copying and copying across the Python/C boundary. As with memory allocations,
\systemname{} triggers a signal (this time, \texttt{SIGPROF}) after a threshold
number of bytes has been copied. It also uses the same temporary file
approach to avoid the problem of lost signals. The
current \texttt{memcpy} sampling rate is set at a multiple
of the allocation sampling rate (currently $2\times$). The ratio of
the of copy sampling and the allocation sampling rates typically has a
proportional impact on the number of interrupts. Since copying is
almost always immediately preceded by an allocation of the same size,
and followed by a deallocation, the current setting maintains copy
samples at roughly the same rate as allocation samples.

\begin{table*}[!t]
    \centering
    \begin{tabular}{lccccccccccc}
        \textbf{Scripting} & \textbf{\texttt{malloc}} & \textbf{Monkey}   & \textbf{Thread} & \textbf{Stack}      & \textbf{Opcode} \\
        \textbf{Language}  & \textbf{interposition}   & \textbf{patching} & \textbf{enum.}  & \textbf{inspection} & \textbf{disassembly} \\
        \toprule
        \textbf{Perl}      & \checkmark (1)  & \checkmark    & \small \texttt{threads->list()} & \texttt{Devel::StackTrace} & \texttt{B::Concise} \\
        \textbf{Tcl/Tk}    & \checkmark (2)  & \checkmark    & \emph{not needed}            & \emph{not needed}                & \emph{not needed} \\
        
        \textbf{Python}    & \checkmark (3)  & \checkmark    & \small \texttt{threading.enumerate()}  & \small \texttt{sys.\_current\_frames()} & \small \texttt{dis} \\
        \textbf{Lua}       & \checkmark (4)  & \checkmark    & \emph{not needed}            & \emph{not needed}                & \emph{not needed} \\
        \textbf{PHP}       & \checkmark (5)   & \checkmark   & \emph{not needed}            & \emph{not needed}                & \emph{not needed} \\
        \textbf{R}         & \checkmark       &  \checkmark  & \emph{not needed}            & \small \texttt{sys.call}  & \small \texttt{disassemble}   \\
        \textbf{Ruby}      & \checkmark (6)   &  \checkmark  & \small \texttt{Thread.list}   & \small \texttt{caller}             & \small \texttt{RubyVM::InstructionSequence}   \\
    \end{tabular}
    \caption{\textbf{Feature support needed for scripting-language aware profiling, with corresponding functions/modules, if needed.} While \systemname{} is a Python profiler, it relies on widely available characteristics of scripting language implementations. (1): Perl's default configuration disables its internal allocator (\texttt{-Dusemymalloc=n}). (2): Tcl/Tk's default configuration also disables its internal allocator (\texttt{-DUSE\_TCLALLOC=0}). (3): Python's allocator can be redirected by setting the environment variable \texttt{PYTHONMALLOC=malloc}. (4): Lua's allocator can be changed via the function \texttt{lua\_setallocf()}. (5): PHP's allocator can be redirected by setting the environment variable \texttt{USE\_ZEND\_ALLOC=0}. (6): Ruby invokes \texttt{malloc} for objects larger than 512 bytes, but does not provide a facility for interposing on smaller object allocations. \label{tab:scripting-language-features}}
\end{table*}

    \section{Evaluation}
    \label{sec:evaluation}

We conduct our evaluation on a MacBook Pro (2016), with a 3.3 GHz
dual-core Intel Core i7, and equipped with 16GB of 2133 MHz DDR3
RAM. The Powerbook running MacOS Catalina (version 10.15.4). All C and
C++ code is compiled with clang version 11.0, and we use version 3.6.8
of the Python interpreter.

\subsection{CPU Profiling Overhead}

This section compares the profiling overhead of \systemname{} to the
suite of existing profilers listed in Table~\ref{tab:comparison}. To
tease apart the impact of the \systemname{} runtime library, we
include the results of \systemname{} without the library, which we
refer to as ``\systemname{} (CPU)'' (as it performs CPU profiling
only, although still separating Python and C execution time), from
``\systemname{} (full)'', which includes both memory and copy volume
tracking.  We conservatively choose CPU-intensive applications to
perform these experiments, as these represent the worst-case for
profiling overheads; overheads are likely to be substantially lower in
applications that spend more time in I/O operations.

\paragraph{\textbf{Benchmarks.}}
While there is a standard benchmark suite for Python known as
\texttt{pyperformance}, most of the included benchmarks are
microbenchmarks, running in many cases for less than a second. As
these are too short lived for our purposes, we conduct our evaluation
on one of the longest running benchmarks, \texttt{bm\_mdp}, which
simulates battles in a game and whose core involves topological
sorting. This benchmark takes roughly five seconds. We also use as a
benchmark program an example used as a basis for profiling in a book
on high-performance Python, which we refer to
as \texttt{julia}~\cite[Chapter~2]{gorelick2020high}; this benchmark
computes the Julia set (a fractal) and runs for seven seconds. We
modify the benchmarks slightly by adding \texttt{@profile} decorators,
as these are required by some profilers; we also add code to ignore
the decorators when they are not used. In addition, we had to add a
call to \texttt{system.exit(-1)} to force \texttt{py-spy} to generate
output. We report the average of three consecutive runs.

\begin{figure}[!t]
\begin{subfigure}[t]{0.99\linewidth}
  \includegraphics[width=\textwidth]{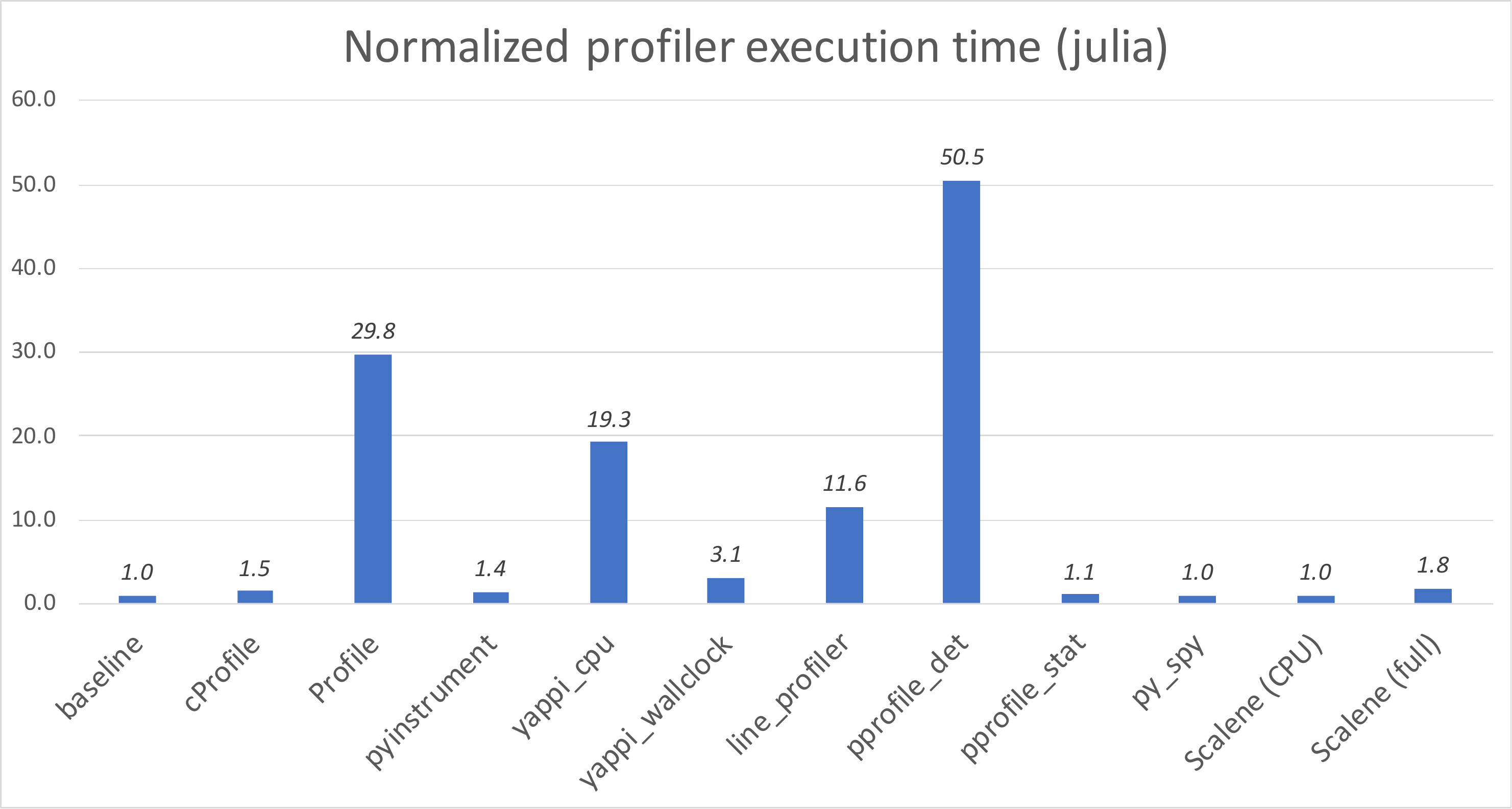} \caption{\textbf{Overhead running the Julia benchmark.}}  \vspace{1em}
\end{subfigure}
\begin{subfigure}[t]{0.99\linewidth}
  \includegraphics[width=\textwidth]{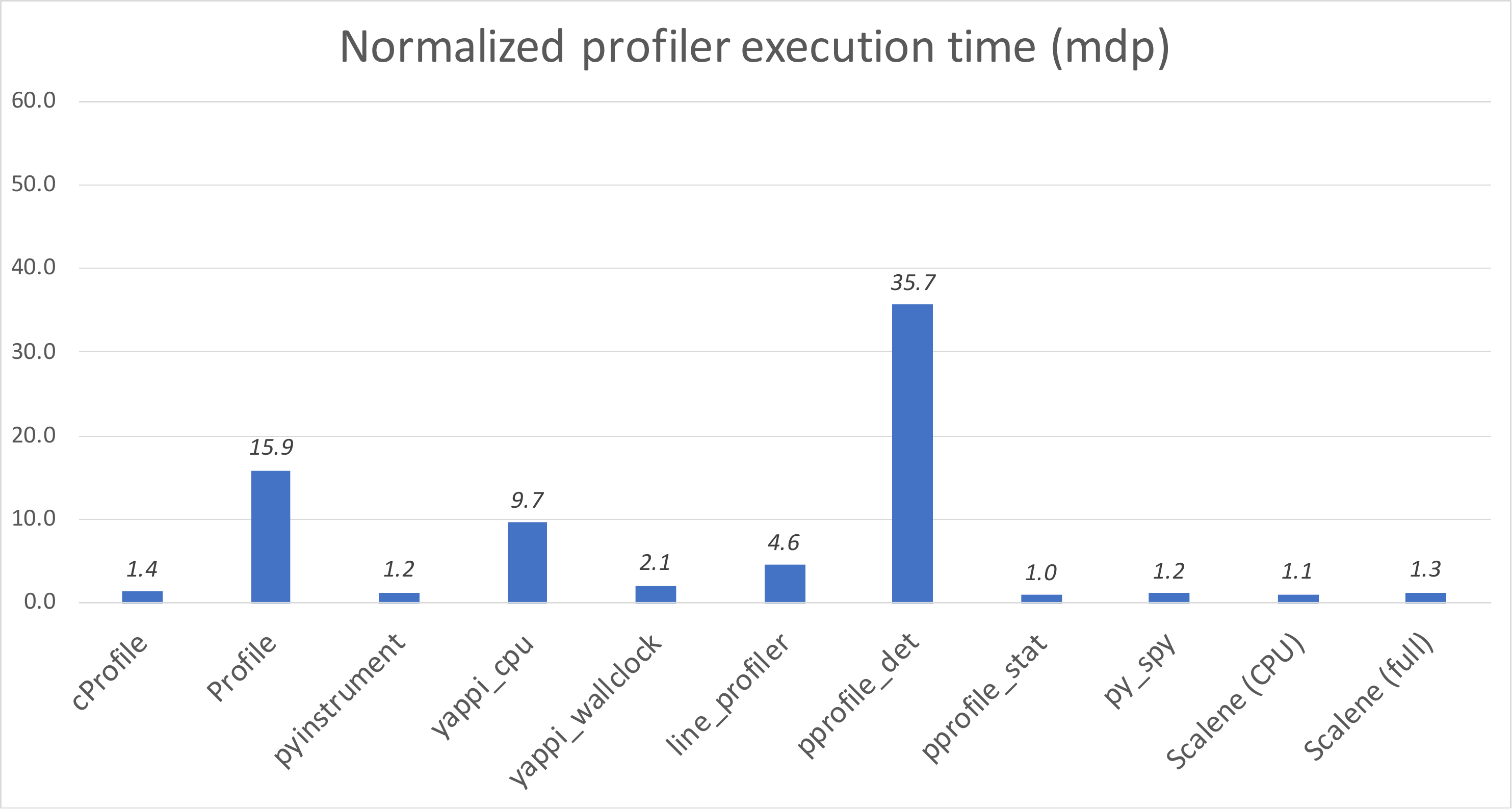}
  \caption{\textbf{Overhead running the mdp benchmark.}}
  \vspace{1em}
\end{subfigure}
\caption{\textbf{Profiling overheads.} Despite collecting far more
  detailed information, \systemname{} is competitive with the
  best-of-breed CPU profilers, imposing no perceivable overhead in its CPU-only version, and
  between 26\%--53\% for its full version.\label{fig:profiler-overheads}}
\end{figure}

Figure~\ref{fig:profiler-overheads} provides these results. In
general,
\systemname{} (CPU only) imposes virtually no overhead, while the full \systemname{}
imposes between 26\% and 53\% overhead.

\subsection{Memory Profiling Overhead}
\label{sec:memory-profiling}

The profilers we examine include just one memory profiler
(\texttt{memory\_profiler}). That profiler's focus is exclusively on
memory profiling; that is, it does not track CPU time at
all. Like \systemname{}, \texttt{memory\_profiler} works at a line
granularity, reporting only average net memory consumption.

We sought to perform an empirical comparison
of \texttt{memory\_profiler}'s performance against \systemname{}. Unfortunately,
\texttt{memory\_profiler} is far too slow to be usable. While it runs for simple examples,
we forcibly aborted it after it had run for at least $100\times$
longer than the baseline; for the Julia benchmark, we allowed it to
run for over 2 hours, but it never completed. In other words, its
slowdown is \emph{at least} $1000\times$. By contrast, \systemname{}
delivers fine-grained memory usage information with vastly lower
overhead.

\subsection{Case Study}
\label{sec:case-study}

\begin{figure*}[!t]
  \begin{subfigure}[t]{0.99\linewidth}
    \includegraphics[width=\textwidth]{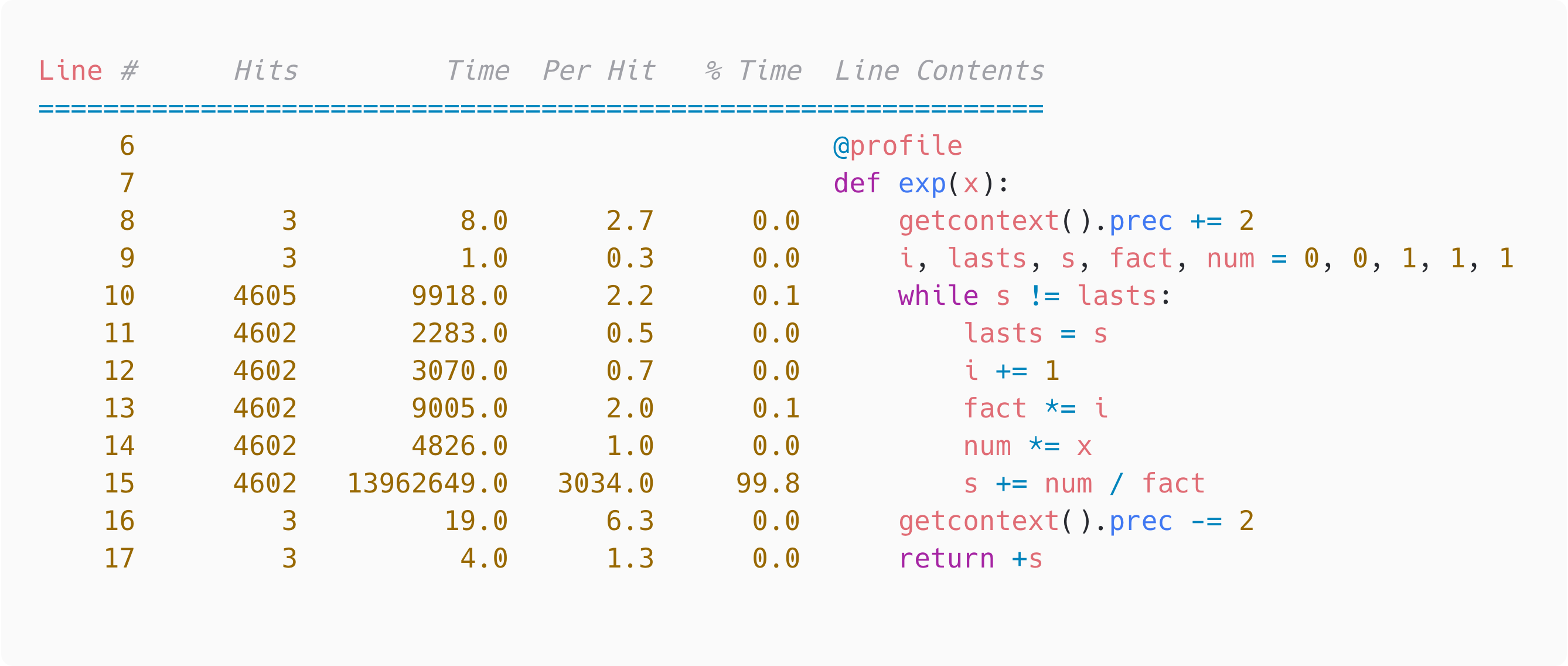}
    \caption{\textbf{Profiling with \texttt{line\_profiler}}. Line 15 is the clear culprit, but the reason is unclear.\label{fig:line-profiler-exp}}
    \vspace{1em}
  \end{subfigure}
  \begin{subfigure}[t]{0.99\linewidth}
    \includegraphics[width=\textwidth]{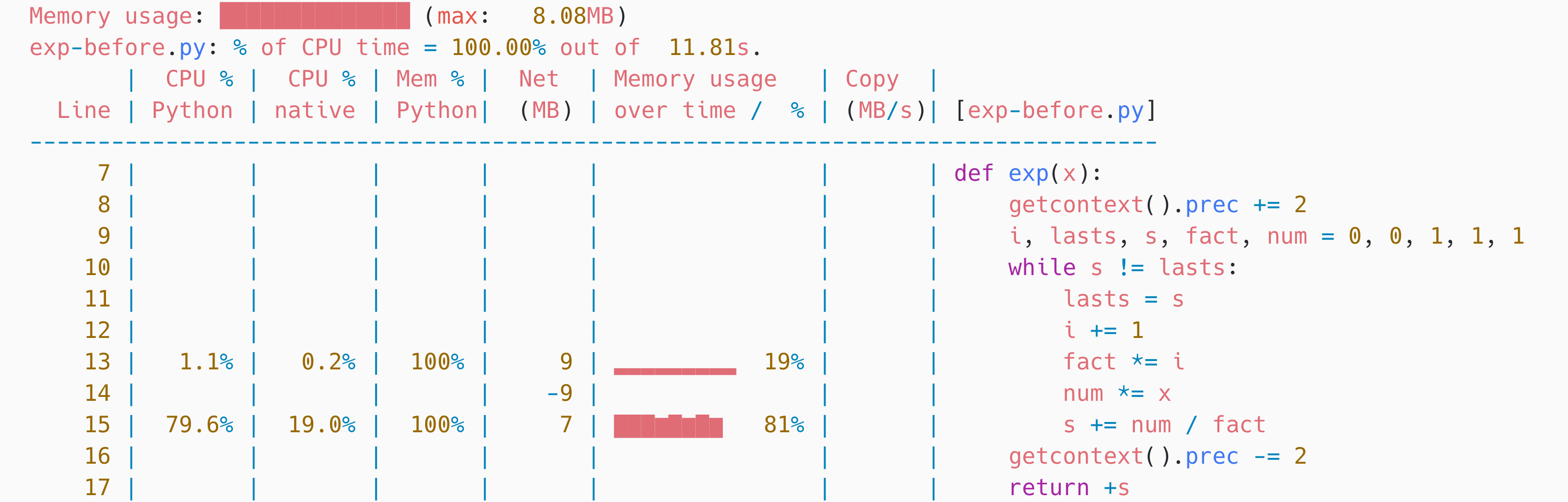}
    \caption{\textbf{Profiling with \systemname{} (before optimization)} \systemname{} reveals that line 15 is allocating and freeing memory at a high rate.\label{fig:scalene-exp}}
    \vspace{1em}
  \end{subfigure}
  \caption{\textbf{Case Study:} This small case study illustrates how \systemname{} can reveal optimization opportunities: in this case, changing a few lines improves performance by over $1,000\times$ ($\S\ref{sec:case-study}$).\label{fig:exp-example}}
  \vspace{1em}
\end{figure*}

In this section, we report how \systemname{} can reveal
previously-unknown optimization opportunities in actual Python
code. This case study is primarily meant to
illustrate \systemname{}'s role in the optimization process, and how
it improves on past work. We note that we do not expect most users
of \systemname{} to identify such enormous optimization opportunities.

We examine code presented in the Python documentation for
the \texttt{Decimal} arbitary-precision library to
compute \texttt{exp} ($e^x$)~\cite{exp-recipe}. Running this code
on \texttt{Decimal(3000)} takes 12 seconds. A standard line-level
profiler (\texttt{line\_profiler}) reports that line 15 is the
bottleneck: computing the ratio
\texttt{num / fact} (Figure~\ref{fig:line-profiler-exp}).
However, \texttt{line\_profiler} does not provide much insight into why this is the case.

When we run \systemname{} on this code, we see an entirely different
story (Figure~\ref{fig:scalene-exp}).
\systemname{} reveals that line 15 is mostly executing in Python, but most importantly,
it shows that it is, somewhat surprisingly, allocating and freeing
objects at a rapid rate. In fact, this single line accounts for 81\%
of the object allocation activity in the program, all in Python. This
fact warranted investigation of the
\texttt{num} and \texttt{fact} variables. Inspecting the values of \texttt{num}
and \texttt{fact} made it clear that both are growing large fast: they
are repeatedly allocating and freeing space for digits.

To address this---that is, to keep the size of these numbers
small---we introduce a variable \texttt{nf} that maintains the
ratio \texttt{num / fact}. This change required the addition of a new
variable, adding one line of code, and deleting two. The result was a drop in
execution time from 12 seconds to 0.01 seconds: an improvement of over
$1,000\times$.

    \section{Related Work}
    \label{sec:related_work}

\subsection{Other Scripting Languages}
\label{sec:other-scripting-languages}

Table~\ref{tab:scripting-languages} provides a breakdown of previous
scripting languages by the characteristics of their standard
implementations. All are dynamically-typed languages, and
their standard implementations are interpreters. This section
describes key features of these scripting languages.

Perl~\cite{DBLP:books/lib/WallS92} was designed by Larry Wall and
first released in 1987. Unusually, Perl does not use bytecodes,
instead using an abstract-syntax tree-based interpreter. It
exclusively relies on a reference-counting garbage collector. Since
Perl 5.8, released in 2002, Perl has provided \emph{interpreter
threads}, which comprise separate interpreters per thread; unlike
traditional threads, all variables and references are thread-local
unless they are explicitly shared~\cite{perlthreads}. Signal delivery is delayed until
interpreter enters a safe state (between opcodes), also since Perl
5.8; previously, it had been signal-unsafe.

Tcl/Tk~\cite{DBLP:conf/usenix/Ousterhout90,wiki:tcl} was designed by
John Ousterhout; its first release was in 1988. It has used a
stack-based bytecode interpreter since version 8.0, released in
1997~\cite{DBLP:conf/tcltk/Lewis96,tcl8.0}, replacing its original
string-based interpreter. It relies exclusively on a
reference-counting garbage collector. Like Perl, Tcl implements a
variant of interpreter threads (as an extension)~\cite{tcl-threads}, with explicit sharing of
variables possible via special operators,\cite[Chapter~21]{welch2003practical}. Unlike
other scripting languages discussed here, core Tcl has no built-in
support for signals since version 8.6, though it is available in
extensions~\cite{tclsignals}.

Python~\cite{DBLP:conf/tools/Rossum97} was designed by Guido van
Rossum and initially released in 1990. It is a stack-based bytecode
interpreter. It has a reference-counting garbage collector, but there
is also an optional \texttt{gc} module that performs mark-sweep
garbage collection. Only one thread at a time can execute in the
interpreter, which is protected by a global lock (the global
interpreter lock, a.k.a., ``the GIL''). Signals are delivered only to
the main thread, and delayed until VM regains control.

Lua~\cite{DBLP:journals/spe/IerusalimschyFF96,DBLP:conf/hopl/IerusalimschyFF07}
was designed by Roberto Ierusalimschy; its first release was in
1993. Lua's interpreter is register-based, rather than
bytecode-based. Lua has never had reference-counting, relying on
stop-the-world mark-sweep garbage collection until incremental GC was
added in version 5.1, released in 2006. Lua had no threads of any kind
until version 4.1; it now has cooperative (non-preemptive)
threads. Signals are delayed until the VM regains control.

PHP~\cite{wiki:php} was designed by Rasmus Lerdorf and first released
in 1994. Its interpreter is similar to a register-based bytecode
(three-address based). It uses a reference-counting garbage
collection, but added a backup cycle collector based on Bacon and
Rajan's synchronous cycle collection
algorithm~\cite{DBLP:conf/ecoop/BaconR01} in PHP 5.3, released in
2009. PHP's default configuration is NTS (Not Thread Safe); threading
can be enabled at build time by turning on ZTS (Zend Thread Safety).
Since PHP 7.0, signal delivery has been delayed until the interpreter
regains control; unlike other scripting languages, PHP delays
delivering signals not just after executing one opcode but only once
the VM reaches a jump or calls instruction.


R~\cite{doi:10.1080/10618600.1996.10474713} was designed by Ross Ihaka
and Robert Gentleman; its first release was in 1995. R is a
reimplementation of the S programming language, developed in 1976 by
John Chambers~\cite{the-new-s-language}, with the addition of lexical
scoping. R has both an AST-based interpreter
and a bytecode interpreter (since version 2.13, released in
2011)~\cite{bytecode-compiler-for-r}. Since its creation, R has
employed a mark-sweep-compact garbage collector.  R is single-threaded
and has no support for signal handling.

Finally, Ruby~\cite{DBLP:books/daglib/0015648} was designed by
Yukihiro Matsumoto (a.k.a., ``Matz'') and first released in
1995. Originally an abstract-syntax-tree based interpreter, it
switched to using a stack-based bytecode interpreter
(``YARV''~\cite{10.1145/1094855.1094912}) with Ruby
1.9~\cite{wiki:rubymri}, released in 2007. Initially, like Lua, it
employed a stop-the-world, mark-sweep garbage collector; generational
collection was introduced in version 2.0, and incremental garbage
collection as of version 2.1. Like Python, Ruby has multiple threads
but these are serialized behind a global-interpreter lock. Signals are
only delivered to the main thread, and they are queued until the
interpreter regains control.

\subsection{Existing Python Profilers}
\label{sec:python-profilers}

Table~\ref{tab:comparison} provides a high-level overview of the
features of all of the major Python profilers of which we are aware.
All but one are CPU profilers.  These profilers fall into two
categories: function-granularity and line-granularity. Most are less
efficient than \systemname{} (particularly in its CPU-only mode),
notably those that rely on Python's built-in support for profiling
(the \texttt{setprofile} and \texttt{setttrace} calls from
the \texttt{sys} and \texttt{threading} modules). Some fail to record
information accurately for multi-threaded applications. None perform
scripting-aware profiling.


Two of the profilers operate in different
modes. Like \systemname{}, \texttt{yappi} can perform either CPU-time
or wall-clock profiling. However, \texttt{yappi}'s CPU-time profiling
mode does not use sampling, making it inefficient, degrading
performance by $10\times$--$20\times$. The wall-clock version is
considerably more efficient, though it still imposes performance
penalties ranging from $2\times$--$3\times$.
Like \texttt{yappi}, \texttt{pprofile} has two different versions: one
is deterministic, relying on instrumentation, while the other uses
sampling. The sampling version imposes low overhead, but the
deterministic version imposes the highest performance penalties of any
CPU profiler we study: from $30\times$--$50\times$.

\subsection{Profilers for Other Scripting Languages}

Like previous Python profilers, profilers for other scripting
languages are essentially variants of traditional profilers for
systems languages; none are scripting-language aware.

Next to Python, Ruby is the scripting language with the most profilers
in wide use. Rbspy is an efficient sampling-based profiler that
inspired the development of Py-Spy~\cite{rbspy}. Another profiler for
Ruby, stackprof, optionally performs object sampling after every so
many allocations~\cite{stackprof}. Unlike \systemname{}, this sampling
does not integrate with CPU sampling, nor does it perform any
scripting-language aware profiling such as separate CPU/memory
attribution, tracking memory usage over time or reporting copy
volume. Finally, Ruby also has a MemoryProfiler that precisely tracks
object allocations at the line granularity, imposing considerable
overheads (up to $5\times$)~\cite{MemoryProfiler}. Like stackprof,
MemoryProfiler cannot simultaneously perform CPU profiling and memory
allocation tracking.

R's standard profiler is Rprof, a line-granularity sampling-based
profiler for CPU and memory consumption; it does not measure CPU time
or memory consumed by native code in libraries. Andersen et al.\
describe \emph{feature-specific profiling}~\cite{andersen2018feature},
a profiling approach that focuses on attributing costs to specific
language features, such as pattern matching or dynamic dispatch. They
present an implementation of this profiler for R that uses Rprof's
sampler. Most feature-specific profiling they describe is orthogonal
and complementary to scripting-language aware profiling. One use case
they describe---identifying when R's copy-on-write policy fails,
resulting in deep copies---would be subsumed by \systemname{}'s copy
volume profiling. A previous R profiler, lineprof, also reports the
number of vector duplications.


Profilers for other scripting languages are 
conventional. Profilers for Lua include
LuaProfiler~\cite{luaprofiler}, LuaTrace~\cite{luatrace}, and
Pro-Fi~\cite{profi}, all function granularity CPU
profilers. Similarly, the standard Tcl profiler is also a
function-level profiler. Perl has a variety of profilers,
include \texttt{Devel::DProf} (also function granularity),
\texttt{Devel::SmallProf} (line granularity), \texttt{Devel::FastProf} (a faster variant of \texttt{Devel::SmallProf} written in C);
the most sophisticated profiler for Perl is \texttt{Devel::NYTProf},
which performs profiling at the file, function, ``block'', and line
granularity~\cite{nytprof}.

    \section{Conclusion}
    \label{sec:conclusion}

This paper introduces scripting-aware profiling, and presents a
prototype scripting-aware profiler for Python
called \systemname{}. \systemname{} both sidesteps and exploits
characteristics of Python---and typical of most scripting
languages---to enable it to deliver actionable information to Python
developers. Its pervasive use of sampling coupled with its runtime
system allow it to capture detailed information with relatively modest
overhead. \systemname{} has been released as open source
at \url{https://github.com/emeryberger/scalene}.

  {\small
  \bibliographystyle{abbrv}
  \bibliography{emery,scalene}
  }
  
\end{document}